\documentclass[sigplan,10pt,screen]{acmart}
\pdfoutput=1

\newif\ifarxiv
\arxivtrue
\ifdefined\arxivbuild
  \arxivtrue
\fi
\ifarxiv
  \newcommand{\figsqueeze}[1]{}
\else
  \newcommand{\figsqueeze}[1]{\vspace{#1}}
\fi

\renewcommand\footnotetextcopyrightpermission[1]{}
\setcopyright{none}
\ifarxiv
  \renewcommand{\shortauthors}{McMahon et al.}
\else
  \acmSubmissionID{899}
\fi

\usepackage{hyperref}
\usepackage{hyperxmp}
\usepackage{xspace}
\usepackage{multirow}
\usepackage{comment}
\usepackage{xfrac}
\usepackage{fancybox, fancyvrb, calc}
\usepackage[center, tight, small]{subfigure}
\usepackage[inline]{enumitem}
\usepackage{amsmath}
\usepackage{blkarray, bigstrut}
\usepackage[font=small]{caption}
\usepackage{url}
\usepackage{graphicx}
\usepackage{pbox}
\usepackage{algorithm, algorithmicx}
\usepackage[noend]{algpseudocode}
\usepackage{balance}
\usepackage{wrapfig}
\usepackage{tikz, pgfplots}
\usetikzlibrary{patterns,matrix,positioning,shadows,backgrounds,arrows,calc,fit,automata,shapes.geometric,arrows,decorations.pathreplacing,decorations.markings,shadings,shapes.symbols,arrows.meta}
\usepackage[capitalise]{cleveref}
\usepackage{tabularx}
\usepackage[T1]{fontenc}
\usepackage[utf8]{inputenc}
\usepackage[normalem]{ulem}
\usepackage{wasysym}
\usepackage{enumitem}
\usepackage{sidecap}
\usepackage{caption}
\usepackage{pifont}
\usepackage[mathcal]{euscript}
\usepackage{setspace}
\usepackage{listings}
\usepackage{footnote}
\makesavenoteenv{tabular}
\makesavenoteenv{table}
\usepackage{soul}
\hypersetup{
  colorlinks=true,      % false: boxed links; true: colored links
  linkcolor=blue,       % color of internal links
  citecolor=magenta,    % color of links to bibliography
  filecolor=cyan,       % color of file links
  urlcolor=red          % color of external links
}
\definecolor{listinggray}{gray}{0.9}
\definecolor{lbcolor}{rgb}{0.9,0.9,0.9}
\graphicspath{{fig/}}

\usepackage{etoolbox}
\usepackage{booktabs}

\usepackage{enumitem}
\setitemize{noitemsep,topsep=0pt,parsep=0pt,partopsep=0pt}
\setenumerate{noitemsep,topsep=0pt,parsep=0pt,partopsep=0pt,itemindent=0em,leftmargin=1.5em}
\setdescription{itemsep=0pt,topsep=0pt,parsep=0pt,partopsep=0pt,itemindent=0em,leftmargin=0.5em}

\usepackage{xargs}                     
\usepackage[colorinlistoftodos,prependcaption,textsize=tiny]{todonotes}
\newcommandx{\change}[2][1=]{}
\newcommandx{\fixed}[2][1=]{}

\definecolor{boxclr}{gray}{0.9}

\pgfdeclarelayer{bg}    % declare background layer
\pgfsetlayers{bg,main}  % set the order of the layers (main is the standard layer)

\makeatletter
\newcommand{\thickhline}{%
    \noalign {\ifnum 0=`}\fi \hrule height 0.8pt
    \futurelet \reserved@a \@xhline
}
\newcolumntype{"}{@{\vrule width 0.8pt}}
\newcolumntype{[}{@{\vrule width 0.8pt\hskip\tabcolsep}}
\newcolumntype{]}{@{\hskip\tabcolsep\vrule width 0.8pt}}
\newcolumntype{!}{@{\hskip\tabcolsep\vrule width 0.8pt\hskip\tabcolsep}}
\makeatother

\newcommand{\cppsnippet}[1]{%
  \begin{lstlisting}[gobble=4]
    #1
  \end{lstlisting}
}

\tikzset{
    position/.style args={#1:#2 from #3}{
        at=(#3.#1), anchor=#1+180, shift=(#1:#2)
    }
}

\tikzset{
  half fill/.style 2 args={fill=#2, path picture={
    \fill[#1, sharp corners] (path picture bounding box.west) --
                         (path picture bounding box.east) --
                         (path picture bounding box.south east) --
                         (path picture bounding box.south west) -- cycle;}},
}

\tikzset{
  nil fill/.style 2 args={fill=#2, path picture={
    \fill[#1, sharp corners] (path picture bounding box.155) --
                         (path picture bounding box.25) --
                         (path picture bounding box.south east) --
                         (path picture bounding box.south west) -- cycle;}},
}

\tikzset{
  almost fill/.style 2 args={fill=#2, path picture={
    \fill[#1, sharp corners] (path picture bounding box.205) --
                         (path picture bounding box.335) --
                         (path picture bounding box.south east) --
                         (path picture bounding box.south west) -- cycle;}},
}

\definecolor{oldcolor}{HTML}{c66541}

\algdef{SE}[DOWHILE]{Do}{doWhile}{\algorithmicdo}[1]{\algorithmicwhile\ #1}%

\newcommand{\cut}[1]{}

\newcommand{\paragraphb}[1]{\vspace{0.075in}\noindent{\bf #1.}}

\colorlet{soulgreen}{green!30}

\colorlet{soulred}{blue!20}

\colorlet{soulblue}{blue!30}

\tikzset{ 
table/.style={
  matrix of nodes,
  row sep=-\pgflinewidth,
  column sep=-\pgflinewidth,
  nodes={rectangle,thick,draw=black,text width={},align=center,font=\small},
  text depth=0.25ex,
  text height=1.25ex,
  nodes in empty cells
},
map/.style={
  matrix of nodes,
  row sep=-\pgflinewidth,
  column sep=-\pgflinewidth,
  nodes={rectangle,draw=black,text width=5em,align=center,font=\small},
  text depth=0.25ex,
  text height=1.25ex,
  nodes in empty cells
},
bigmap/.style={
  matrix of nodes,
  row sep=-\pgflinewidth,
  column sep=-\pgflinewidth,
  nodes={rectangle,draw=black,text width=26em,align=center,font=\small},
  text depth=0.25ex,
  text height=1.25ex,
  nodes in empty cells
},
memcell/.style={
  draw, 
  very thick, 
  text width=0.25em, 
  text height=0.25em
},
}

\tikzstyle{startstop} = [rectangle, rounded corners, minimum width=3em, minimum height=1em,text centered, draw=black, fill=red!30]
\tikzstyle{io} = [trapezium, trapezium left angle=70, trapezium right angle=120, minimum width=2.5em, minimum height=1em, text centered, draw=black, fill=blue!30]
\tikzstyle{process} = [rectangle, minimum width=1.5em, minimum height=1em, align=center, draw=black, fill=gray!30]
\tikzstyle{decision} = [diamond, minimum width=3em, minimum height=1em, align=center, draw=black, fill=SkyBlue!30]
\tikzstyle{arrow} = [thick,->,>=stealth]
\tikzstyle{monolog} = [fill=SkyBlue!30]

\tikzset{
  on each segment/.style={
    decorate,
    decoration={
      show path construction,
      moveto code={},
      lineto code={
        \path [#1]
        (\tikzinputsegmentfirst) -- (\tikzinputsegmentlast);
      },
      curveto code={
        \path [#1] (\tikzinputsegmentfirst)
        .. controls
        (\tikzinputsegmentsupporta) and (\tikzinputsegmentsupportb)
        ..
        (\tikzinputsegmentlast);
      },
      closepath code={
        \path [#1]
        (\tikzinputsegmentfirst) -- (\tikzinputsegmentlast);
      },
    },
  },
  mid arrow/.style={postaction={decorate,decoration={
        markings,
        mark=at position .5 with {\arrow[#1]{stealth}}
      }}},
}

\algnewcommand{\IIf}[1]{\State\algorithmicif\ #1\ \algorithmicthen}%
\algnewcommand{\EndIIf}{\unskip\ }%
\algdef{SE}[DOWHILE]{Do}{doWhile}{\algorithmicdo}[1]{\algorithmicwhile\ #1}%
\algnewcommand\algorithmicforeach{\textbf{for each}}%
\algdef{S}[FOR]{ForEach}[1]{\algorithmicforeach\ #1\ \algorithmicdo}%

\newcounter{mynote}[section]

\definecolor{brightgreen}{rgb}{0.0, 0.5, 0.0}

\def\ie{{i.e.}}
\def\eg{{\em e.g.}\xspace}

\def\namex{Callosal}
\def\name{\namex\xspace}
\def\fs{\name{}FS\xspace}

\def\inode{\texttt{inode}\xspace}
\def\page{\texttt{page}\xspace}
\def\as{\texttt{address\_space}\xspace}
\def\writeop{\texttt{write()}\xspace}
\def\readop{\texttt{read()}\xspace}
\graphicspath{{fig/}{fig/results/}}

\begin{document}
\title{\name: Rethinking the Kernel Page Cache \\for Disaggregated Storage and Memory}

\ifarxiv
\author{James McMahon}
\affiliation{%
  \institution{University of Utah}
  \country{}}

\author{Amit Samanta}
\affiliation{%
  \institution{University of Utah}
  \country{}}

\author{Seung-seob Lee}
\affiliation{%
  \institution{Yale University}
  \country{}}

\author{Ryan Stutsman}
\affiliation{%
  \institution{University of Utah}
  \country{}}

\author{Anurag Khandelwal}
\affiliation{%
  \institution{Yale University}
  \country{}}
\else
%\author{Paper \#}
%\affiliation{%
%   \institution{}
%   \country{}}
\fi
% acmart requires to define abstract before maketitle
% !TEX root = ../paper.tex
\begin{abstract}

Classic distributed filesystems like NFS centralize filesystem operations on a server and use custom, filesystem-specific approaches for managing cache consistency across client nodes. This limits scalability, and it leads to weak consistency, 
ruling out applications that concurrently access and mutate shared files.
Our key insight is that these legacy designs are now too storage-centric; with today's large memories, OS page caches do most of the ``heavy lifting'' in accessing data, not the filesystem and block storage layers.

In response to this, we present \name, 
a page cache for a cluster of machines that is centered on shared disaggregated memory.
Rather than serializing operations through a filesystem server and using ad hoc caching like NFS, \name{} is built around the existing Linux page cache paths and is accelerated by RDMA and programmable switches. 
\name{} delivers the strong consistency that programs expect with local filesystems (like in ext4 and tmpfs) while avoiding the additional network access delays and bottlenecks that host-based solutions incur.
We compare \name{} against NFS, CephFS, and local filesystems; often \name{} can provide caching performance similar to local filesystems while also scaling across nodes.
For mixed read/write workloads, on average \name{} has 4.6$\times$ higher throughput than NFS when contention is not extreme, and it
offers strong consistency guarantees like CephFS, while outperforming it.
For a file-backed key-value store with 1\% writes and uniform access, \name{}
scales 1.7$\times$ across 4 nodes, while CephFS collapses across nodes.

\end{abstract}

 \maketitle
\pagestyle{plain}
% !TEX root = ../paper.tex
\section{Introduction}
\label{sec:intro}

Cloud systems disaggregate storage from compute to improve
utilization, elasticity, and capacity sharing
~\cite{racksched, shu2023disaggregated, vuppalapati2020building,
min2021gimbal, klimovic2016flash, RedyVLDB2022}. Fast networks make remote
block storage practical for elastic cloud volumes
~\cite{nvmeof,i10,hwang2021rearchitecting}, where a remote block device can
be attached, migrated, or remapped. However, this architecture does
not keep file caches coherent across compute nodes, since it does
not order the cached reads and writes that applications rely on when multiple
machines cache and mutate the same files.
The absence of a coherent file cache complicates applications that maintain
shared mutable state in files. For example, a distributed, file-backed key-value store could
divide ownership of keys across workers while still allowing any worker to read
keys updated by remote workers. Without coherence, such an
application must build its own
solution.

\begin{figure}[t] 
\centering
\includegraphics[scale = .6]{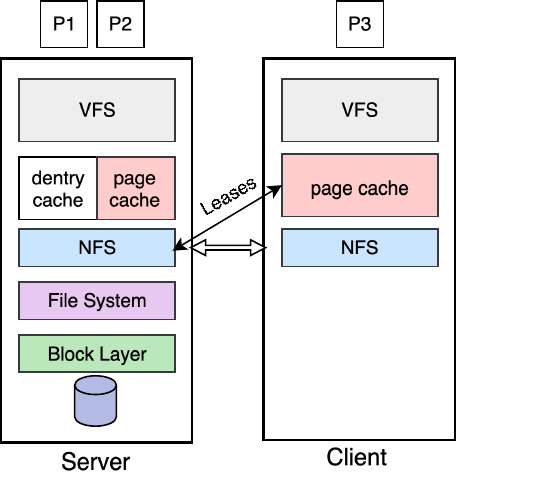}
\vspace{-5mm}
\caption{NFS centralizes filesystem operations and uses ad hoc cache coordination to provide weak guarantees.}
\label{nfs}
\vspace{-6mm}
\end{figure}
In recent years, 
growing
DRAM capacities mean that the kernel
page cache absorbs the vast majority of file reads and writes, even for
workloads backed by remote storage. 
In Linux, 
data is first brought into the page cache and
then accessed, modified, and synchronized there.
Hence, for many
workloads, 
the page cache is
the primary focus of performance and consistency for I/O, not filesystems themselves.

However, distributed storage systems continue to treat the
filesystem or storage layer as the central point of coordination. Traditional
networked filesystems such as NFS (Figure~\ref{nfs}) serialize operations
through a server and rely on ad hoc client caching~\cite{nfs}, resulting in
weak consistency guarantees for cached file data. This can improve performance
when clients mostly operate independently, but it is not sufficient for
applications that require reads to observe completed writes from other machines.
For example, the Linux \texttt{open()} manpage states:
\begin{quote}
\small
\texttt{O\_APPEND  may lead to corrupted files on NFS\\
filesystems if more than one process appends\\data to a file at once.%
}
\end{quote}
\noindent
More recent systems like CephFS provide stronger, POSIX-like
semantics~\cite{ceph,ceph-consistency}, but do so by introducing substantial
coordination overheads that limit performance when files are concurrently
accessed and modified. 
In either approach, workloads that mix reads and writes to shared files suffer,
either via frequent, high-cost application-level coordination of caches or via
weakened consistency.

The key insight of this paper is that distributed file access can be made both
\textbf{efficient} and \textbf{strongly consistent} by using RDMA and
in-network coherence to directly coordinate the kernel page cache across
machines. Rather than treating each node's page cache as private and optimizing
separately, we argue that cached file-backed pages should be treated as shared,
first-class objects whose coherence is enforced explicitly. This aligns
coordination with the dominant execution path of file I/O and avoids the need
for heavyweight filesystem-level protocols. Tightly integrating distributed
coherence into the page cache affords other scalability optimizations and
simplifications. For example, safely coordinating writes to shared block
storage is simplified by building on the page ownership invariants enforced by
distributed coherence, admitting a simple filesystem design that avoids
centralization.

\begin{figure}[t] 
\centering
\includegraphics[scale = .6]{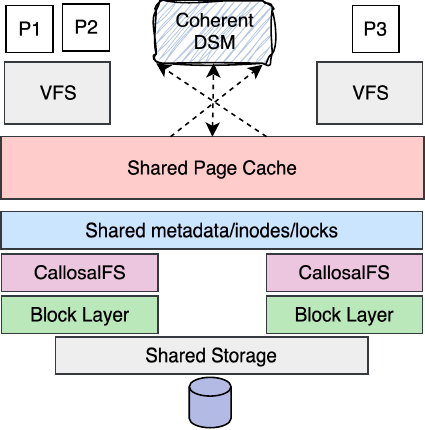}
\vspace{-3mm}
\caption{\name{} creates a shared page cache in Linux, building atop a network-accelerated DSM.}
\label{sysx}
\vspace{-6mm}
\end{figure}

In this paper, we present \name{}, which extends the Linux kernel page cache
with \emph{cross-machine} coherence. \name{} allows nodes
to cache, read, and modify file-backed pages independently while maintaining
real-time ordering through page-granular coordination (Figure~\ref{sysx}). To
achieve this, \name{} builds on recent advances in network-accelerated
distributed shared memory (DSM) using RDMA and an in-network coherence
directory to efficiently track page ownership and propagate invalidations
across nodes~\cite{mind}. 
\name{} interposes at the page cache layer rather than rearchitecting the
filesystem or block storage stack. It integrates with the Linux VFS and preserves
the interface that the page cache exposes to filesystems, allowing unmodified
applications to access shared files using standard POSIX system calls. Most
read and write operations are satisfied locally from cached pages, while
coordination is invoked only when nodes concurrently access overlapping regions
of a file. As a result, nodes can operate independently in the common case
while still providing strong consistency when sharing occurs.

This page-cache-centric design contrasts with conventional distributed
filesystems that centralize ordering or metadata management. By using DSM and
shifting distributed coordination directly to the page cache, \name{} avoids
single-node bottlenecks and reduces unnecessary communication for disjoint
accesses. 
However, designing a page cache that leverages DSM requires solving several
challenges. Foremost, directly moving VFS page cache data structures into DSM
does not work, since many generic kernel data structures would need to be moved
to DSM, which would require sweeping changes to the kernel and its existing
filesystem structures (\eg, to move \page{} structs, \texttt{address\_space}
structs, etc.). Even if this approach worked, it would not perform well;
contention on updating file metadata fields, long lock acquisition times, and
false sharing would render it ineffective. To be efficient, it is essential
that remote DSM operations only be invoked when necessary (for consistency) and
when the semantics of metadata operations cannot be exploited to avoid remote
communication. This leads to a design in \name{} that ensures DSM operations
are generally not triggered when a node reads or writes cached data locally.
\name{} also uses DSM in a stylized fashion to synchronize inodes on updates
efficiently, and, in cases where coordination is necessary, it uses its
in-network coherence directory to target communication precisely to active
sharers at line-rate without additional network delays or server-based
bottlenecks.

Overall, this paper makes the following contributions:
\begin{enumerate}[leftmargin=1em,itemindent=0pt]

\item 
\textbf{Page-cache-centric coherence abstraction.} We present a design
    that extends the Linux kernel page cache into a coherent, cross-machine
    abstraction using network-accelerated DSM.

\item
\textbf{A decentralized design for strongly consistent shared file data.} We
    build on DSM coherence invariants to coordinate file-backed pages and
    metadata without centralizing operations, enabling
    transparent, strongly consistent access to shared files coordinated via
    RDMA and an in-network coherence directory.

\item
\textbf{A practical Linux integration.} We implement this design in the Linux
    kernel as \name{} and evaluate it through \fs{}, a prototype filesystem
    supported by \name{}'s coherent page cache, showing
    how to interpose at the page cache layer while preserving VFS interfaces
    and standard file operations.

\end{enumerate}

We evaluate \name{} and \fs{} across four nodes connected by a Barefoot Tofino
switch running the MIND coherence protocol. On controlled page-sharing
benchmarks, \name{} matches or exceeds NFS for a read-heavy workload with
stronger consistency guarantees than NFS. For a mixed
read/write sharing workload, \name{} outperforms NFS by 4.6$\times$ on average for the
DRWOM benchmark with 5\% writes. Against CephFS, which provides stronger
filesystem semantics than NFS, \name{} improves throughput by an order of
magnitude on multiple workloads. Finally, on a file-backed key-value store where workers share access to mutable state,
\fs{} demonstrates the benefit of coherent file caching by providing correct behavior unlike NFS and dramatically higher throughput than CephFS.

% !TEX root = ../paper.tex
\section{Background and Motivation}
\label{sec:motivation}

This section reviews the distributed filesystem designs summarized in
Table~\ref{tab:tab2}, explains why the page cache is the right layer for
coherence, and summarizes the MIND substrate that \name{} builds on.

\subsection{Distributed Filesystem Designs}
\label{sec:dfs-designs}
Distributed filesystems allow applications to scale beyond a single machine in
capacity and IOPS~\cite{patil2011scale}. NFS~\cite{nfs} is widely used, but its
centralized design limits scalability: a single server handles client requests
and can become a bottleneck as client counts and request rates increase.
NFS also provides weak close-to-open consistency. It ensures that when
a client closes a file, a future open of the same file will see that client's
changes, but it does not guarantee that reads observe concurrent updates in
real-time order~\cite{linearizability}. This is weaker than the behavior
applications expect from typical Linux filesystems, and it can lead to stale
reads and data inconsistencies.

CephFS is a distributed filesystem that provides strong consistency using
leases to manage client caches~\cite{ceph-consistency}. CephFS is built on top
of Ceph, which includes a scale-out distributed object store that
separates operations on data from metadata management~\cite{ceph}; file data is
stored and accessed through object store servers. 
While CephFS's design should scale in principle, our evaluation shows
throughput collapse in shared-file workloads with writes.
In our observations, it also transfers data between clients and OSDs
during read-only workloads, and its read throughput does not scale linearly
across threads (Figure~\ref{drbl}, \S\ref{sec:evaluation}).

Assise is a distributed filesystem that uses client-local, byte-addressable
non-volatile memory (NVM) for fast file access~\cite{assise}.
Clients access and update state from userspace, and updates are replicated from
userspace using RDMA.
Assise fails to meet our transparency requirement, since it
requires using a specialized Assise-specific library for file access, and
its generality is limited by NVM hardware.

\subsection{Page Cache in Modern Filesystems}

Page caches are central to Linux filesystem performance because they avoid
repeated requests to slower storage~\cite{tricache, scalacache, streamcache}.
For networked filesystems, the page cache also avoids network traversals, which
can be slower than local storage access. Linux filesystem operations are built
around cached pages: if a file page is not cached, it is first added to the page
cache before the operation proceeds. Therefore, synchronization for operations
that affect common file pages is handled primarily in the page cache.

The page cache is where NFS and CephFS make opposite consistency/performance
tradeoffs for shared file data. NFS allows clients to cache file data without
keeping cached pages coherent with concurrent updates at other clients, reducing
latency and coherence traffic at the cost of weaker consistency. CephFS
also caches data to avoid remote operations, but it uses remote coordination to
keep cached data consistent, increasing the cost of operations that require
cache validation or invalidation.

This page-cache tradeoff also appears in local filesystems. Local filesystems
generally do not fully serialize concurrent reads and writes: avoiding strict
serialization keeps the read path fast, but reads can observe the effects of
partially completed writes. Enforcing serialization would require reads to
acquire a lock on the cached page, slowing one of the most performance-critical
filesystem operations. This shows that filesystem semantics and
performance already depend on how aggressively the page cache synchronizes
cached file data. A key hypothesis of \name{} is that distributed filesystems can
preserve strong shared-file semantics without giving up scalable read
performance by synchronizing cached file pages directly.

\begin{table}[t]
  \centering
  \scalebox{1.0}{
    \small
    \begin{tabular}{l|ccc}
      \toprule
      \multicolumn{1}{c|}{\textbf{Filesystems}}
      & \multicolumn{1}{c}{\textbf{Consistency}}
      & \multicolumn{1}{c}{\shortstack{\textbf{In‐Kernel/}\\\textbf{Transparent}}}
      & \multicolumn{1}{c}{\shortstack{\textbf{Efficient}\\\textbf{Sharing}}} \\
      \midrule
      NFS~\cite{nfs}       & Weak   & Yes & Yes \\
      CephFS~\cite{ceph}   & Strong & Yes & No  \\
      Assise~\cite{assise}& Strong & No  & Yes \\
      \textbf{\name}       & \textbf{Strong} & \textbf{Yes} & \textbf{Yes} \\
      \bottomrule
    \end{tabular}
  }
  \vspace{1mm}
  \caption{Comparison of distributed filesystems}
  \label{tab:tab2}
  \vspace{-10mm}
\end{table}

\subsection{Cache-coherent Shared Memory in MIND}
\label{sec:mind}

MIND is a rack-scale DSM system that places memory management logic within the
network fabric, providing scalable and elastic memory sharing across compute
nodes~\cite{mind}. MIND connects compute and memory nodes via a programmable
network switch. Each compute node uses a small amount of DRAM as a cache. If an
application accesses a page-granularity cache line that is not present in the
DRAM cache, it triggers a page fault. The fault handler initiates coherence
transactions via cache controller logic in the kernel. The kernel sends requests
for the cache line with read (\ie, shareable) or write (\ie, exclusive)
permissions for faulting memory load or store operations.
The
programmable switch implements the cache directory, issuing invalidations to
relevant compute nodes and updating cache line permissions and sharer state as
necessary. Invalidation
requests are handled by in-kernel cache controller logic at compute nodes,
which removes the cache lines from the cache and sends an acknowledgment to the
permission-requesting compute node via the programmable switch. If none of the
compute nodes have the cache line in their cache, the data is fetched from
disaggregated memory. \name{} employs MIND to realize its distributed page
cache. MIND has additional virtual-memory components that are unnecessary for
understanding \name{}.

% !TEX root = ../paper.tex
\section{\name{} Overview}
\label{sec:overview}

We begin by outlining \name's key properties and how it achieves them:

\vspace{1mm}
\begin{description}
\item[Transparency.]
By integrating directly into a kernel filesystem's page cache, there is no need for application-level changes, userspace libraries, or special API calls. Applications access a \name{} filesystem
using conventional \texttt{read} and \texttt{write} system calls as with local files.

\vspace{1mm}
\item[Efficient Strong Consistency.]
Nodes using a \name{} page cache use an MSI-style invalidation-based protocol to ensure coherence for the pages they operate on; hence, nodes work independently unless they access overlapping file regions.
For efficiency, \name{} handles synchronization and invalidations via an in-network coherence directory.
Unlike NFS, \name{} maintains the strong consistency guarantees expected of a local filesystem's page cache.

\vspace{1mm}
\item[Scalability.]
\name{}'s design avoids any centralized node for file state, coherence, or storage management. Many past distributed filesystems have allowed client caching while centralizing storage updates (NFS) or have decentralized storage while centralizing metadata and cache consistency management (CephFS). 
\name{} uses its coherence- and page-cache-centric design to avoid both bottlenecks. Nodes work independently on pages, allowing them to handle page write-back. This enables nodes to share block storage while ensuring safety and consistency.

\vspace{1mm}
\item[Low Complexity.]
\name limits kernel changes via hooks into Linux's page cache, and it preserves most of the generic Linux page cache code paths and interfaces. The \name{}-enhanced page cache works as the standard local page cache for conventional filesystems while enabling remote coherence for \name{}-enhanced filesystems.
\end{description}

\paragraphb{\name{} design choices}
Traditional distributed filesystem designs are storage-device-centric, where a single node orders operations to storage devices or to individual pages~\cite{nfs,ceph}. Clients in these systems typically use specialized protocols to ensure consistency with the underlying block device rather than relying on the standard Linux page cache logic.

\name{} is a new page cache approach that inverts this design. It is motivated by design trends in single-node filesystems, where larger DRAM sizes in modern servers enable most read and write operations to be handled by the page cache rather than the block layer. With consistent advances in low-latency networking (\eg, via RDMA~\cite{infiniswap, kalia2016design, aifm}) and in-network processing, \name aims to show that efficient DSM (\S\ref{sec:mind}) enables page-cache-centric distributed filesystems that are both simple and efficient.

In \name, no one node acts as the sole ordering entity for operations on block devices. Instead, \name{} uses its DSM-based page cache approach~\cite{mind, cc-primer} to enforce consistency. Specifically, it relies on the Single-Writer-Multiple-Reader (SWMR) invariant, enabled by MIND's scalable in-network coherence, to ensure that file operations are correctly ordered. Avoiding communications with a centralized server node reduces network delays and avoids bottlenecks that would emerge if a host needed to broadcast invalidations.

While \name{} builds over a DSM (\S\ref{sec:mind}), it cannot directly replace the Linux page cache with the DSM since its memory consistency semantics do not directly map to those of a Linux filesystem. Instead, \name{} preserves Linux's generic page cache abstraction at each node and uses the DSM's shared memory to synchronize filesystem data across page caches on different nodes. This leaves the page cache intact while avoiding DSM and remote access overheads for most operations since the Linux page cache at each node can handle them locally.

Akin to single-node filesystems, \name{}'s inverted design refocuses filesystem complexity back to the page cache, enabling the remote block layer to remain modular and simple.
Hence, \name{}'s design is compatible with a variety of remote block stores. For example, nodes using \name{} 
could in principle share access to an NVMe-oF or Amazon EBS volume~\cite{nvmeof,nvmeof_2016} or other state-of-the-art disaggregated block storage~\cite{i10,hwang2021rearchitecting}.
Operations that make it past the page cache are already ordered (via the DSM's SWMR semantics), and this behavior can help to enforce synchronization 
of reads and writes to 
the block layer. 

\subsection{Operation Overview}

Nodes using \name{} mount and access a filesystem with a single shared 
disaggregated page cache.
Unlike typical networked filesystems, with \name{}, page caches on different nodes maintain a consistent view of pages through a shared virtual address space enabled by MIND. This ensures that reads and writes to each file's pages are coherent across different nodes' page caches without requiring a central node.

When a process issues a read, \name maps the file offset to a page. A MIND coherence request is issued if the page exists in \name's global page cache (\ie, in MIND's address space) but is absent in the node's local page cache. The MIND directory (located at the network switch) forwards the request to the node whose page cache has a copy, which responds with the page to the requesting node.
The node caches it in read-only mode and completes the file read. If no node's local page cache contains the page, the page must be fetched from storage and placed in the shared address space.

Writes are handled similarly: before writing to a page in MIND, the node ensures it has a writable copy of the page. MIND's coherence protocol ensures that conflicting, concurrent writes to a page on other nodes are serialized.

Compared to a local filesystem like ext4, a filesystem using \name{} requires two main components to be added/modified. First, all nodes sharing the filesystem map the shared address space provided by MIND. Second, while file reads and writes in \name{} files are resolved and synchronized through the page cache in the coherent address space, the filesystem needs metadata~\cite{filemetadata, tablefs, ganger1994metadata} (\eg, inodes) to track where persistent copies of blocks are stored. Like most other filesystems, this is implemented as a kernel module, which we call \fs.  \fs stores some of its metadata in MIND's shared address space so nodes can access and modify a shared view of it.

\subsection{Consistency Model}
\label{sec:consistency-model}

\name{} provides strong consistency semantics. Every \readop{} or \writeop{} that starts after a remote or local \writeop{} completes is guaranteed to see the effects of that earlier \writeop{}. That is, all operations respect {\em real-time ordering}~\cite{decentralized-storage,linearizability}.

Like the page cache in local Linux filesystems, \name{} is not linearizable~\cite{linearizability}, which would also require preserving the appearance of atomic \readop{}s and \writeop{}s. For example, in ext4, an ongoing \readop{} can interleave with an ongoing \writeop{}, with the effect observed on some bytes but not others.
\name{} has the same semantics for concurrent \readop{}s and \writeop{}s.
This is because it preserves the generic page cache behavior for local operations. Concurrent remote operations can exhibit similar interleaving effects, but only at the page granularity, providing semantics as strong as ext4's local semantics.
One small exception is that in \name{}, even \writeop{}s can interleave at page granularity if they are concurrent on different nodes for higher performance; in ext4 \writeop{}s always hold an \texttt{inode} lock, preventing interleaving between \writeop{}s.

\begin{figure}[t]
%\vspace{-10mm}
\centering
\includegraphics[width=0.95\columnwidth]{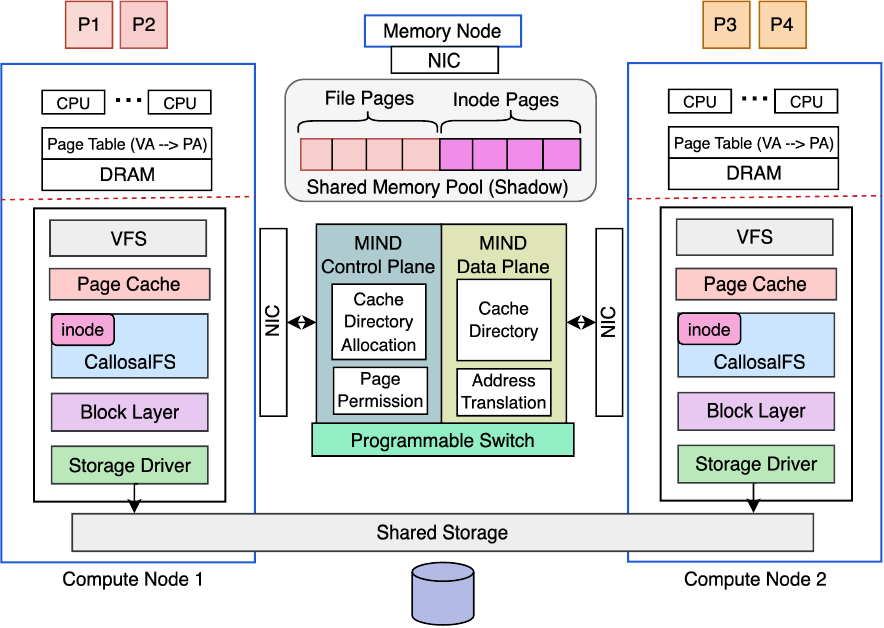}
\caption{\name Overview}
\label{fig:callosalfs_design}
\vspace{-5mm}
\end{figure}

\section{\name{} Design}
\label{sec:design}

We first illustrate \name{}'s design (Figure~\ref{fig:callosalfs_design}) by walking through how reads (\S\ref{ssec:reads}) and
writes (\S\ref{ssec:writes}, \S\ref{ssec:invalid}) work over the coherent page cache.
We then discuss the necessary filesystem metadata fields for correct 
interaction with \name{}'s page cache. 
Using \name{}'s page cache leaves the node's local VFS and page cache layers largely unchanged from 
the generic implementation when introducing the adjustments to support synchronization.
In Linux's page cache, each file (\inode{}) has an \as{} that maps file offsets to \page{}s, each of which refers to a page buffer. Each \inode{} and \page{} object has metadata fields and fields for synchronizing access to it. For each \inode{} and each \page{} associated with a file block, \name{} adds a new lock (a \texttt{rwsem}) that we call the \textit{remote lock} for that \inode{} or \page{}. These locks use the shared remote address space for synchronization. They are used, for example, to prevent nodes from invalidating the local page copy during ongoing local \readop{} and \writeop{} operations. Since the added remote locks are closely integrated with existing local locks, \name{} piggybacks on the local page cache's deadlock-freedom guarantees.
Our modified page cache is compatible with any standard, unmodified filesystem. \name{}'s changes direct synchronization operations on \inode{}s to the underlying filesystem code, which allows the filesystem to synchronize correctly. Sometimes, this results in local locking; other times, it requires synchronization across nodes. This defaults to the generic behavior provided for other filesystems.

\begin{figure}[t]
\centering
\subfigure[Read Path]{%
\includegraphics[scale = .52]{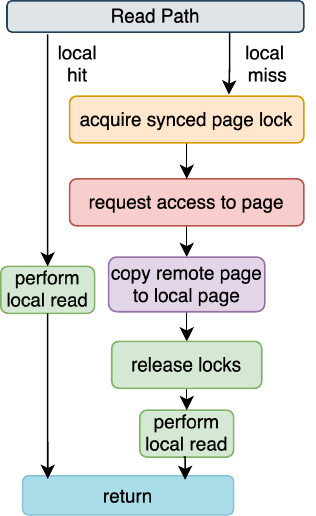}
\label{fig:reads}}
\subfigure[Invalidation Path]{%
\includegraphics[scale = .52]{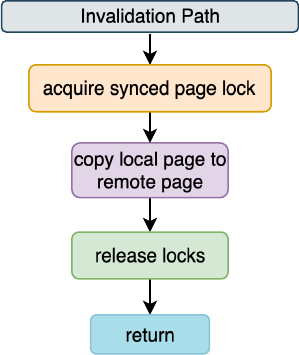}
\label{fig:invalidations}}
\vspace{-3mm}
\caption{Block diagrams of \name{}'s read and invalidation paths.}
\vspace{-5mm}
\end{figure}

\subsection{File Reads}
\label{ssec:reads}

As we will outline in this subsection, file reads in \name{} retain Linux's standard behavior, though they diverge in some cases while still remaining POSIX compliant.

In Linux, reads that hit in the page cache optimistically skip acquiring locks.
\name{} behaves similarly; \readop{}s to an up-to-date page that the node's page cache has in read-only or writable modes are served locally and without acquiring page locks (Figure~\ref{fig:reads} left side).
This means that, as with ext4,
\readop{}s that access a locally cached copy of a page can overlap with a local \writeop{} to a page, so partial local writes can be observed.

For \readop{}s that access pages not cached locally (Figure~\ref{fig:reads} right side), \name{} acquires the synced page locks (i.e., remote and local locks)
page-by-page for each read page, which limits this interleaving to page granularity.
After locking, the reading node fetches it in read-only mode from MIND, which blocks until a remote copy of the page is returned. The request for the page is transparently forwarded to the current owner of the page by MIND's in-network cache directory. If the owner has the page in its page cache with writable permission, the remote copy of the page is invalidated.
Ultimately,
MIND's SWMR invariant in \name{} helps ensure that all \readop{}s that are issued after a \writeop{} completes in real time will observe its effects.

This page-granular synchronization with remote writers introduces a subtle consistency challenge: ensuring that nodes observe all completed writes in the same total order~\cite{linearizability}; we describe this issue and our solution in \S\ref{sec:race}.

\subsection{File Writes}
\label{ssec:writes}

\begin{figure}[t] 
\centering
\includegraphics[width=0.56\columnwidth]{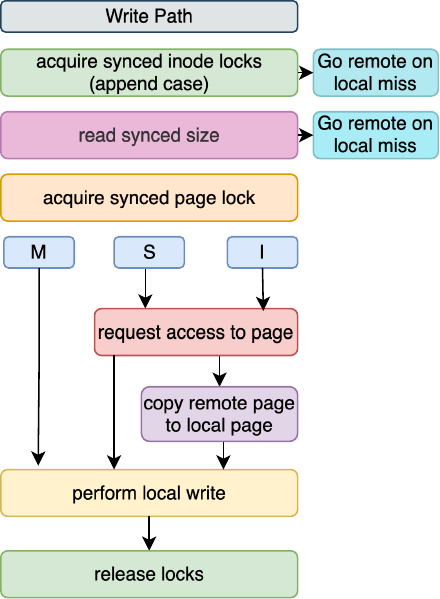}
\vspace{-2mm}
\caption{Block diagram of \name{}'s write path.}
\label{fig:writes}
\vspace{-6mm}
\end{figure}

Similarly, Linux's generic \writeop{} path is largely preserved. One key change is that \name{} allows concurrent writes from different nodes to different file pages as long as they do not change the file size. In the simplest case (Figure~\ref{fig:writes}, left path), a copy of the page from MIND is local on the node and in writable mode (denoted by \texttt{M}). In this case, the writer acquires the synced page locks. This is where \name{} relies on MIND's SWMR coherence invariant; for pages in writable mode that are present locally, acquiring these locks is a local operation that merely prevents invalidation before the locks are released. By acquiring these locks, the write path also serializes concurrent writes to the page.

If the page is either uncached (\texttt{I}) or locally cached in read-only mode (\texttt{S}), in addition to the synchronized page lock, the node must upgrade its permission by requesting the current page owner over the network and waiting for a response. MIND handles this request (at the network switch) by ensuring all remote copies of the page have been invalidated. It also provides the response, which includes an acknowledgment of the permission upgrade and the page content if the page was originally uncached (\texttt{I}); if it was in read-only mode (\texttt{S}), the response contains only the acknowledgment. Subsequently, the write is served locally.
As a consequence of synchronizing the locks, sizes, and page data, appends work intuitively. Since appends extend the file size atomically across concurrent writes and other appends, appended data is never interleaved with concurrent writes, even if it spans multiple pages.

\subsection{Page Invalidation}
\label{ssec:invalid}

Using the DSM approach for coherence allows \name{} to read state without triggering remote communication in many common cases, thereby achieving significantly lower read latency. For example, if a page is already available locally, then the remote lock for the page is never acquired. Generally, this would lead to non-real-time-ordered behavior since stale pages might be read, but \name{} relies on MIND's coherence invalidation (Figure~\ref{fig:invalidations}) messages to avoid this.

Whenever a node requests a page in writable mode to handle a \writeop{}, all copies on other nodes must be acknowledged as invalidated before the \writeop{} proceeds.
Since \writeop{}s on each node acquire and hold the remote lock for a page while they perform invalidation, all of a \writeop{}'s effects on each page are globally linearized via MIND's SWMR property.

\readop{}s that start after a \writeop{} to a page always process invalidations first, ensuring that the local cached copy is up-to-date before proceeding.

\subsection{In-network Coherence Directory for RDMA}
Unlike traditional distributed filesystems that typically serialize coordination at a centralized metadata server to maintain a coherent view, \name relies on MIND's in-network cache directory for that purpose. 
This design provides the following benefits: 

\begin{enumerate}[leftmargin=1em,itemindent=0pt]
\vspace{1mm}
\item 
Read and write operations are naturally ordered by the switch pipeline at line rate as they traverse the network, thereby avoiding synchronization overhead from a centralized coordinator.
\vspace{1mm}
\item
Compute nodes are relieved from tracking the serialization point or the locations of cached pages; the switch automatically resolves data locations and generates invalidation requests, thereby eliminating extra network roundtrips for metadata lookup and reducing traffic by targeting only the relevant sharers. 
\vspace{1mm}
\item
Maintaining the coherence directory in the switch simplifies fault recovery, as data ownership can be reconstructed from the switch's state, which itself can be protected via control-plane replication and data-plane reconstruction. 
\vspace{1mm}
\item
Finally, it improves scalability by eliminating hotspots due to a centralized coordinator, since the switch can process coherence operations at line rate~\cite{mind}.
\end{enumerate}
\vspace{1mm}

\subsection{Metadata Synchronization and Durability}

\name{} handles \readop{}s and \writeop{}s, including appends and truncations, ensuring it handles the hot path operations for most applications. While much filesystem metadata is only accessed on \texttt{open()}, \texttt{close()}, and \texttt{stat()} calls, some crucial file metadata must be synchronized to support basic operations as well.
We use \name{}'s coherent shared memory in a specialized way to efficiently support metadata synchronization.

Specifically, in addition to a local copy, each \inode{} has a corresponding shadow \inode{} stored in MIND that tracks an up-to-date view of its important fields. Critically, this includes a file's size and its remote \inode{} lock. They are synchronized frequently since \writeop{}s can implicitly extend the file size and \readop{}s must validate against it.

\begin{figure}[t]
\centering
\includegraphics[width=0.74\columnwidth]{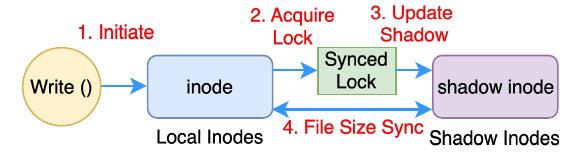}
\vspace{-5mm}
\caption{Overview of inode synchronization during writes.}
\label{fig:inode-sync}
\vspace{-4mm}
\end{figure}

During a \writeop{}, the local \inode{}
lock synchronizes local writes, and the \inode{} remote lock synchronizes remote accesses. The \inode{} remote lock (a readers-writer lock like the remote lock on each \page{}) is acquired in read mode on the \inode{} (Figure~\ref{fig:inode-sync}). The lock is acquired in write mode when the file size changes due to a write. This ensures that writes that change the file size serialize correctly. 
Changes to a file's size are updated in the shadow \inode{}.
For wider application support, other metadata fields can be incorporated into the shadow \inode{}s to synchronize them (\eg, \texttt{mtime}), though we currently do not include them.
Pathologically, the accessed time (\texttt{atime}) attribute would force coherence actions on every read if it were synchronized; for this reason, \texttt{atime} is commonly disabled in practice, even on local filesystems.

Other structures like directory entries are never accessed on \readop{} and \writeop{}; hence, they do not impact the applications we have targeted, though they can be supported in a similar fashion on top of \name{}'s coherent memory.
Writeback can rely on MIND's SWMR invariant to ensure that each node can coordinate writeback for a particular dirty page without additional synchronization. \texttt{fsync()} would require coordinating writeback across hosts, which would require a two-phased approach where all nodes perform writeback and then acknowledge completion to the initiator. WekaFS's write cache mount mode offers fsync with similar writebacks~\cite{Weka2020Architecture, wekawriteback}. Writeback passes dirty blocks to the underlying file system in our prototype, but correct distributed \texttt{fsync} is not fully implemented.

\section{Implementation}
\label{sec:impl}
\name{}'s primary goal is to preserve local filesystem page cache performance and 
behavior while replicating that behavior across nodes. It also seeks to make
minimal adjustments to the core Linux kernel by relying on generic kernel code
paths where possible and filesystem-specific callbacks.

\name{} was designed to use MIND, but it deviates significantly from MIND's
original use cases. \name{} uses MIND's cache coherence protocol to quickly and
efficiently synchronize page caches between nodes. The inclusion of MIND is a
significant addition to the filesystem's cache hierarchy because it must bridge
two separate caching mechanisms --- \ie, each node's page cache and MIND's own
coherence protocol --- that depend on each other to function correctly. MIND
serves as a way to synchronize the local caches of nodes instead of
functioning as the cache itself. This takes advantage of the strengths of both
systems.
In this section, we discuss the
implementation of \name's 
synchronized page cache in \fs{}. The \fs{} module builds on top 
of a simple skeleton filesystem 
kernel module~\cite{simplefs}, and it is comprised of about 3,700
lines of C.

\subsection{Kernel Modifications}
Apart from the modifications introduced by MIND, \name{} requires only a small set of changes (250 lines of code) to the core kernel in Linux 4.15~\cite{linux415}. These modifications primarily export previously unexported functions, allowing them to be accessed from kernel modules.

However, \name{} changes some paths in the kernel itself. 
The two main changes affect the (common) inode lock and size read/write code paths since these operations often occur outside of the filesystem, and they are essential to synchronize for \name's correctness. For both cases, wrapper functions are used similarly to existing VFS file system hooks. We add a branch in these functions to allow for a callback into the filesystem, enabling inode locking and size reading to be controlled and synchronized by the filesystem during each access for filesystems that desire interposition.

\subsection{\fs Implementation}
\fs{} is a basic filesystem implementation that incorporates \name{}, providing enough functionality to evaluate \name{}'s design. For example, \fs{} provides a synthetic view of storage where all file pages are initially zeroed.
To support synchronization with remote invalidations, \fs{} tracks the data structures being shared. This is important for cases like truncations, where the address of a page in the page cache may change due to reallocation.
\fs{} synchronizes page data, \inode{} locks, and file size.
\fs uses spinlocks and read-write semaphores to synchronize operations on these structures.

\subsection{MIND Coherence Protocol Modifications}
To implement page and inode coherence, \name{} modifies MIND's coherence protocol in two key ways. First, in MIND, a 64-bit address field represents a compound process identifier (the left-most 16 bits) and virtual address (the remaining 48 bits). In \name, the low-order 48 bits are repurposed to find shadow file and inode pages in MIND, enabling page and inode coherence.
Second, since MIND and its data structures were originally designed for user processes, \name{} uses a 
separate hash table and statically allocated arrays to store the permissions of pages and inodes on each compute node. Page, inode, and metadata synchronization operations, including those on inode size, look up and update these permissions (\S\ref{sec:design}). \name{} also employs additional locking and invalidation queues compared to MIND. This is because I/O operations in \name{} must block other operations, while in MIND, invalidation requests can be handled immediately.

\subsection{Syncing Remote Operations with Local Reads}
\label{sec:race}

Before reading each page, a \readop{} first checks to see if its local page cache contains
an up-to-date copy of the page.  If it does, the copy in the page cache is read. If a page is out of date, the remote sync path is taken (Figure~\ref{fig:reads}), which updates the local page cache from the up-to-date remote copy.
Without care, if these two paths overlap on different threads, it can lead to a subtle consistency bug.

\begin{figure}[t] 
\centering
\includegraphics[width=0.95\columnwidth]{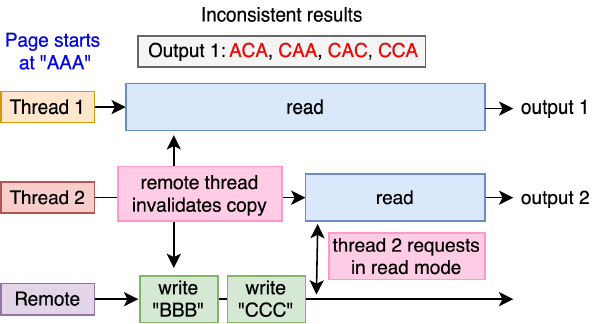}
\vspace{-2mm}
\caption{A consistency problem can arise without validation.}
\label{partial_copy}
\vspace{-6mm}
\end{figure}

Figure~\ref{partial_copy} shows an example. Thread~1 begins a \readop{} on an up-to-date page containing \texttt{AAA}. Subsequently, threads on a remote node invalidate the page and then make two writes to the page (replacing it with \texttt{BBB} and then with \texttt{CCC}). After these writes, if Thread~2 starts a \readop{} that is concurrent with Thread~1's \readop{}, then it fetches \texttt{CCC} and begins writing it into the local copy of the page byte by byte. As Thread~1 copies data out of the page, it may see a mix of the old and new data, for example, \texttt{ACA}, but this interleaving should not be possible according to \name{}'s consistency model.
Specifically, since \texttt{write(BBB)} completes before \texttt{write(CCC)} in this history, if Thread~1 observes a \texttt{C} in the buffer, then no byte later in the buffer should be seen as \texttt{A}.

Fortunately, there is a simple fix with very low cost. 
Each page has an associated version number that is updated before and after a remote read occurs,
and readers validate the version number after copying data out of the page. If the version number changes during a \readop{}, the copy is repeated. In our experiments, this retry triggers exceedingly rarely.

\subsection{Limitations of the Prototype}
\label{sec:discussion}

\name{}'s implementation supports efficient \readop{} and \writeop{} operations including appends. Other operations are implemented to a sufficient degree to support the workloads we run. Complete writeback and durability are unimplemented, though \name{} does pass dirty blocks to \fs{} for writeback.

Currently, the \name{} prototype preallocates or statically divides much of the DSM to assess its read and write performance, without being affected by dynamic memory allocation/reallocation.
Adding a simple memory allocator on top of MIND would address this limitation~\cite{huang2025tigon}. Aspects of \name{}'s per-filesystem implementation are also preallocated or statically allocated for similar reasons.

\section{Evaluation}
\label{sec:evaluation}
We evaluate \name{}'s page cache performance through our test filesystem \fs{}. We compare its performance with existing distributed (NFS and Ceph) and local filesystems (ext4 and tmpfs) in several benchmarks. In our evaluations, we compare the throughput and scalability of each filesystem using workloads that vary in contention and read/write ratios. We also evaluate whether \name{}'s design improves throughput for applications that require strong consistency.

\begin{figure*}[t] 
\centering
\subfigure[DRBL]{%
\includegraphics[scale = .3225]{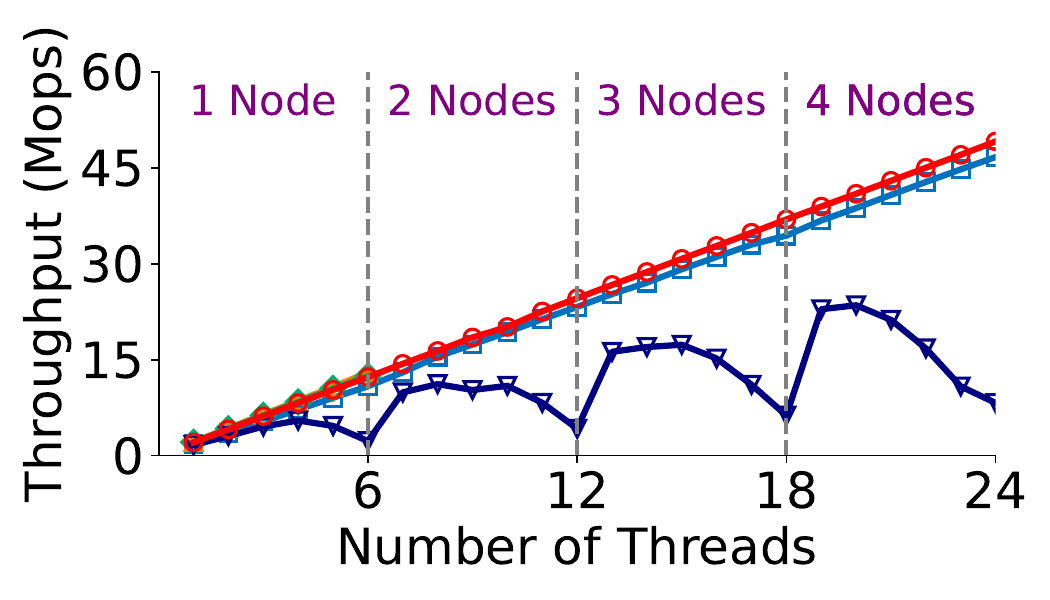}
\label{drbl}}
\subfigure[DRBM]{%
\includegraphics[scale = .3225]{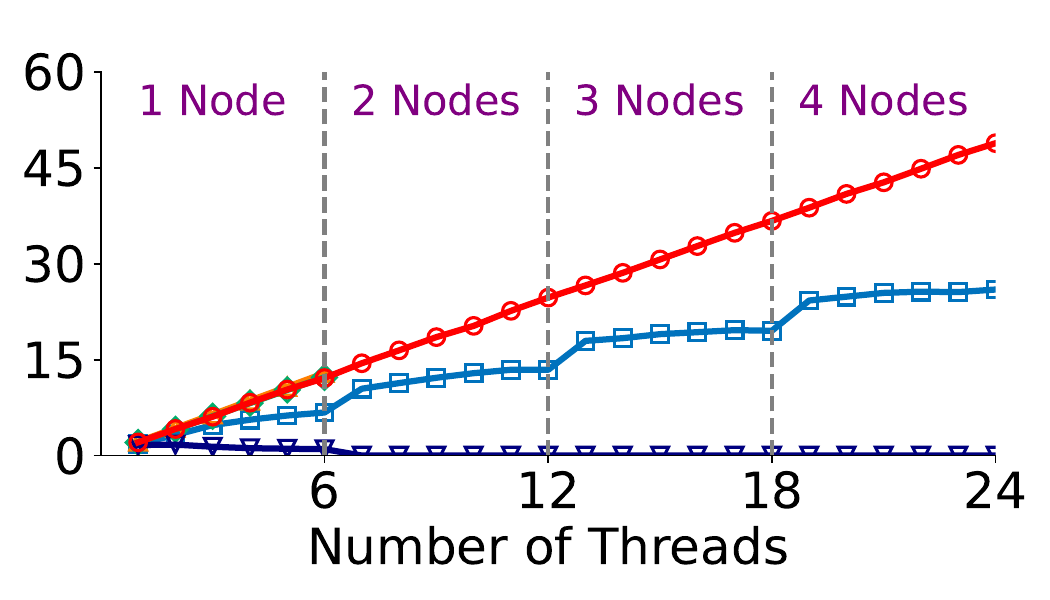}
\label{drbm}}
\subfigure[DRBH]{%
\includegraphics[scale = .3225]{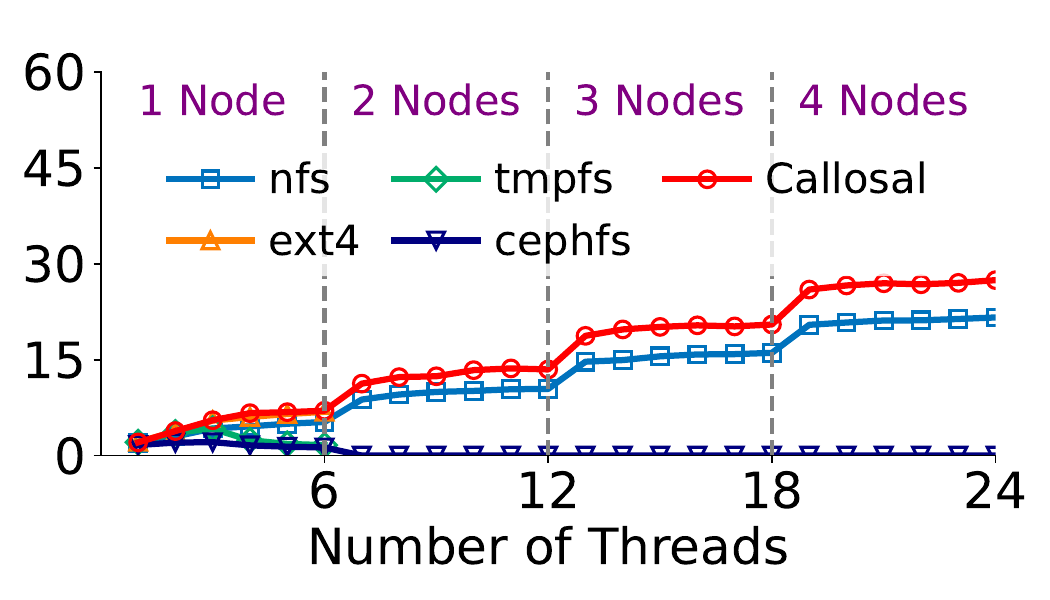}
\label{drbh}}
\vspace{-5mm}
\caption{Read performance of a block in these scenarios. (a) DRBL: reads a block in a private file, (b) DRBM: reads a private block in a shared file, and (c) DRBH: reads a shared block in a shared file. \name{}, ext4, NFS, and tmpfs scale linearly. }
\label{read_fxmark}
\vspace{-6mm}
\end{figure*}

\begin{figure*}[ht!] 
\centering
\subfigure[DWOL]{%
\includegraphics[scale = .3225]{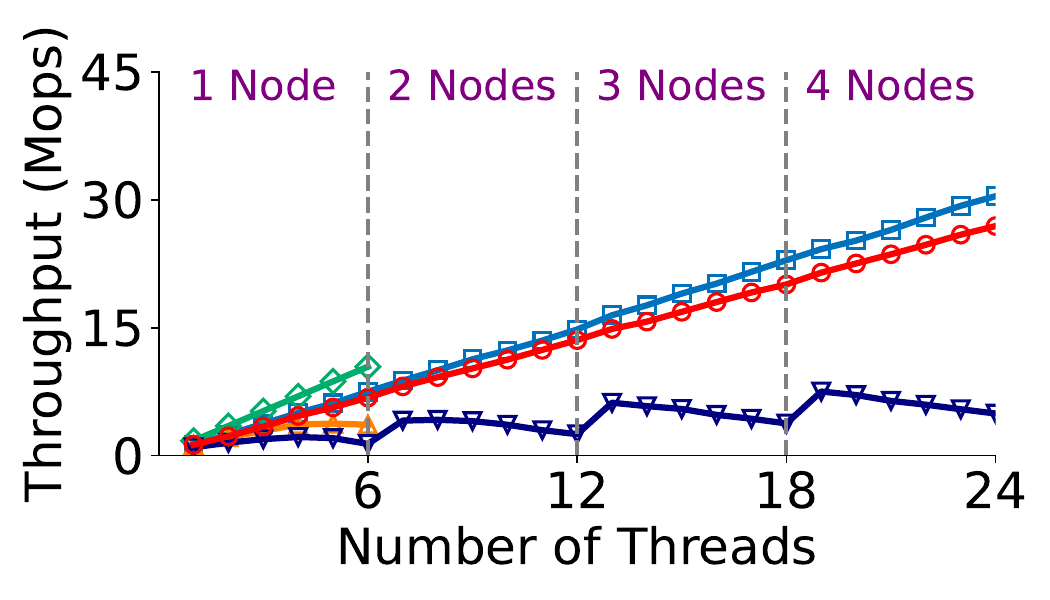}
\label{dwol}}
\subfigure[DWOM]{%
\includegraphics[scale = .3225]{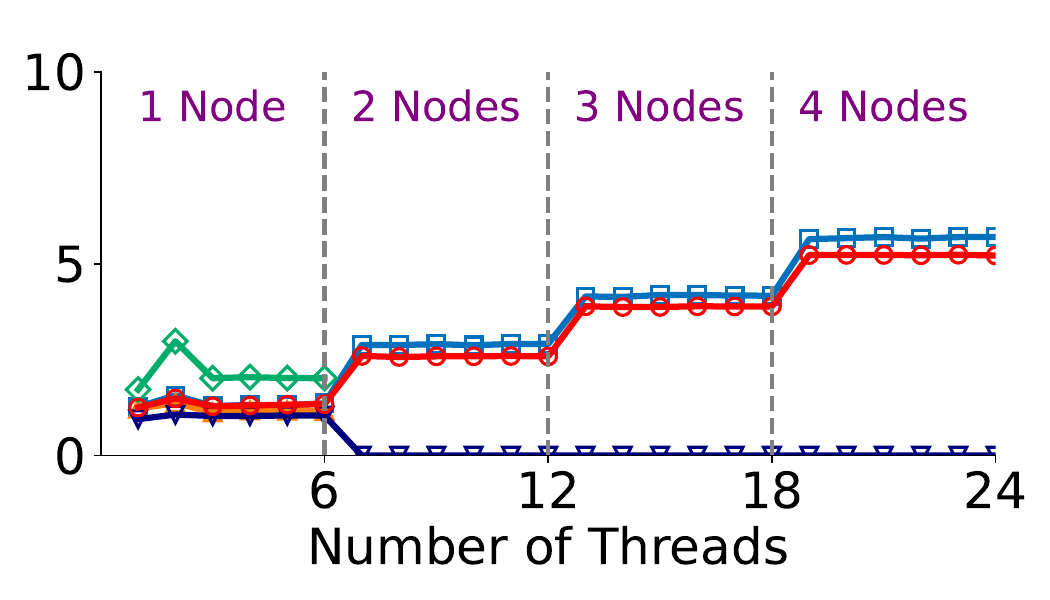}
\label{dwom}}
\subfigure[DWOH]{%
\includegraphics[scale = .3225]{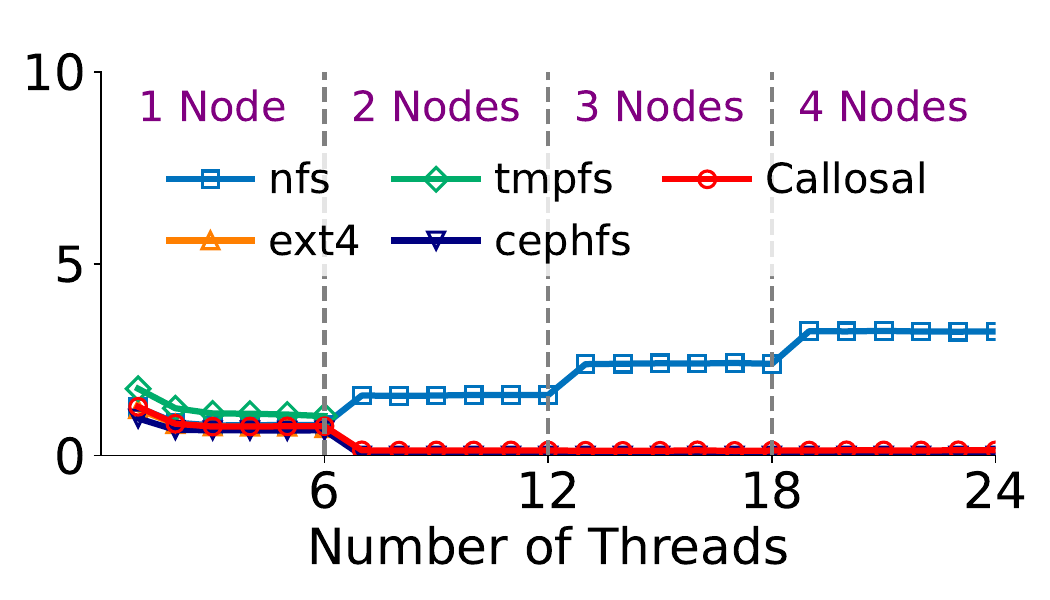}
\label{dwoh}}
\vspace{-5mm}
\caption{Write performance of a block. (a) DWOL: writes a block in a private file, (b) DWOM: writes a private block in a shared file, and (c) DWOH: writes a shared block in a shared file. \name{} delivers lower throughput than NFS under conflicting writes due to strong consistency.
}
\label{}
\vspace{-4mm}
\end{figure*}

\paragraphb{Cluster Setup} 
We evaluate \fs{} on four servers connected by a programmable switch. The switch is a 32-port EdgeCore Wedge featuring a 6.4 Tbps Tofino ASIC and an Intel Broadwell processor, along with 8 GB of RAM and a 128 GB SSD. Each of the four servers has two 12-core Intel Xeon processors and two NVIDIA CX-5 100 Gbps NICs. Our benchmarks use up to six of the 12 cores on each node, leaving the other cores to handle MIND coherence messages. We scale \fs{} by first scaling on the number of threads running on one compute node before introducing the next compute node. As we add compute nodes, we balance the threads across the nodes. This approach yields a fair comparison to local filesystems, while consolidating clients on the fewest number of nodes, thereby reducing contention.

\paragraphb{Baselines} We compare \fs{} with four widely used filesystems: ext4~\cite{ext4}, NFSv4~\cite{nfs}, tmpfs~\cite{tmpfs} and CephFS~\cite{ceph}. Each file system uses its default metadata settings; hence, they all update last accessed time, though \fs{} only tracks \texttt{atime} locally. We ran tests with \texttt{atime} disabled as well, but it had no significant effect on CephFS and NFS performance. NFS exports files from one server to many clients via RPC. ext4 is a commonly used local filesystem for Linux systems. Although ext4 cannot scale to multiple nodes, it serves as a baseline for \fs{}'s local performance on a single node. tmpfs is a local, in-memory filesystem; tmpfs provides no durability and only runs on one node, but it illustrates a practical upper bound on the performance that \fs{} could achieve on a single node. 

CephFS has a scale-out design, provides strong consistency, and it has a kernel-based client. It represents the closest baseline to \fs{}.
We run CephFS on two nodes, distinct from the four that run the benchmarking clients to avoid interference. The two nodes each host a Ceph OSD that serves the objects that make up file contents. One of the nodes also runs Ceph's MDS, which tracks file metadata and file content locations along with monitoring and management daemons. Clients access Ceph using the 100~Gbps network.

We also tested NFS v4 with RDMA transport (NFSoRDMA) for the \texttt{fxmark} benchmarks, but we did not see any performance improvement because NFS largely hits in its local cache, and thus we left it out of further discussions.
While other filesystems provide strong semantics, they are not fair comparisons for \name{} and as a result \fs{}. For example, a common class of filesystems changes the client interfaces (GFS~\cite{gfs}, HDFS~\cite{shvachko2010hadoop}) or has user-level implementations (Assise~\cite{assise}, Octopus~\cite{octopus}). The other common class weakens consistency semantics for performance (NFS~\cite{nfs}, AFS~\cite{afs}, Coda~\cite{coda1990}), which breaks applications that mutate shared data. However, unlike these systems, \fs{} is an in-kernel filesystem that provides strong consistency even when files are concurrently modified on different nodes.

\paragraphb{Benchmarks and Applications}
We evaluate \fs{} using the \texttt{fxmark}~\cite{fxmark} microbenchmarks, a standard set of benchmarks designed to test filesystem scalability. These benchmarks primarily evaluate filesystem reads or writes in isolation while varying contention. We chose the benchmarks to focus on the 
caching performance of each filesystem.
\textbf{DRBL} reads a block in a private file on each thread on each node (low contention reads).
\textbf{DRBM} reads a private block in one file shared by all threads (medium (metadata) contention reads).
\textbf{DRBH} reads one shared block in one shared file (high (page-level) contention reads).
\textbf{DWOL} overwrites a private block in a private file (low contention).
\textbf{DWOM} overwrites a private block in a shared file (medium contention).
\textbf{DWOH} overwrites a shared block in a shared file (high contention).
We also include mixed read/write benchmarks derived from \texttt{fxmark}, a
producer-consumer queue, and a file-backed key-value store.

\subsection{Read Throughput}

\paragraphb{DRBL} In DRBL, every thread reads pages from independent files. \fs{} nearly matches local filesystem performance when run on one node, and it scales linearly across nodes (Figure~\ref{drbl}). As it scales across compute nodes, \fs{}'s performance slightly beats NFS's performance.
Though CephFS also caches at each node, its added overhead prevents it from scaling linearly across nodes.

\paragraphb{DRBM} In DRBM, each thread reads from a private block in a shared file (Figure~\ref{drbm}). All of the filesystems except CephFS scale well. 
As expected, \fs{}'s performance scales and is nearly identical to its performance on DRBL;
NFS's performance is lower. 
Even though each thread reads from a private block, having multiple clients reading from the same file causes CephFS to send extra messages to the metadata server, which causes its throughput to collapse.

\paragraphb{DRBH} Figure~\ref{drbh} shows the results for DRBH in which every thread reads the same block in the same file. tmpfs performance drops as thread count scales. This
is due to it spending a large amount of time on reference counting~\cite{fxmark}.
However, \fs, NFS, and ext4 scale well with increasing thread counts. \fs{} consistently outperforms NFS, both within a single node and as it scales across nodes.
CephFS's performance in DRBH follows a similar trend to DRBM.

\begin{figure*}[t] 
\centering
\quad \hspace{-8mm}
\subfigure[DRWOM 5$\%$ Writes]{%
\includegraphics[scale = .3225]{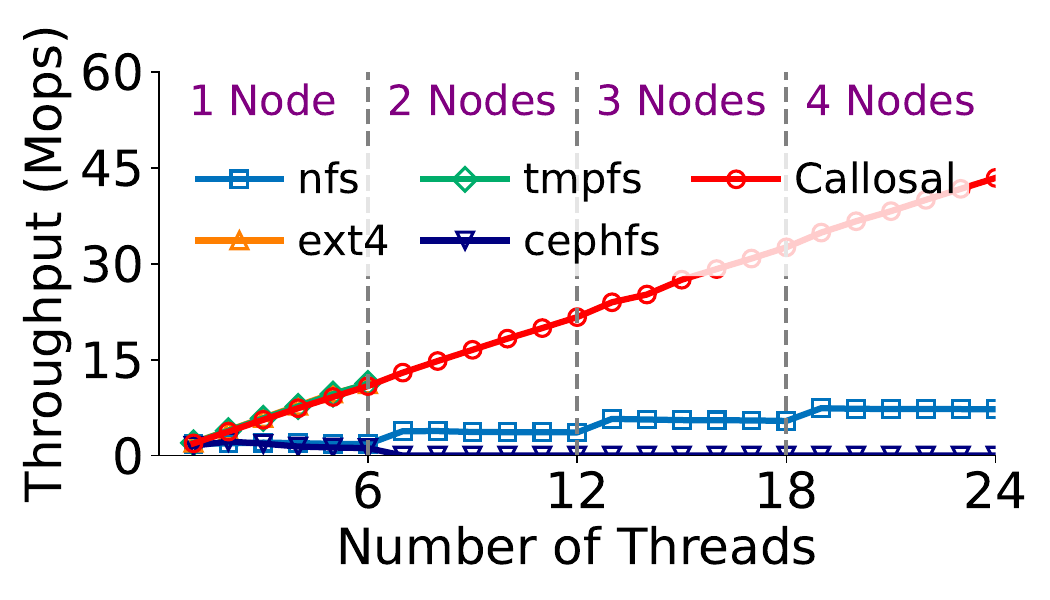}
\label{DRWOM}}
\subfigure[DRWOM]{%
\includegraphics[scale = .3225]{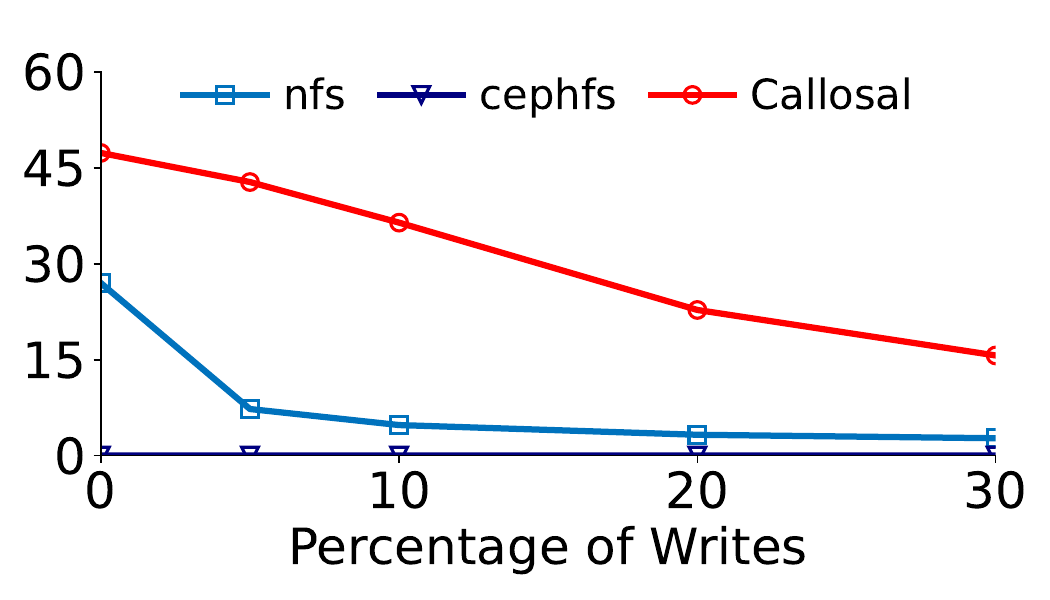}
\label{DRWOM_writes}}
\subfigure[DRWOM Single Node]{%
\includegraphics[scale = .3225]{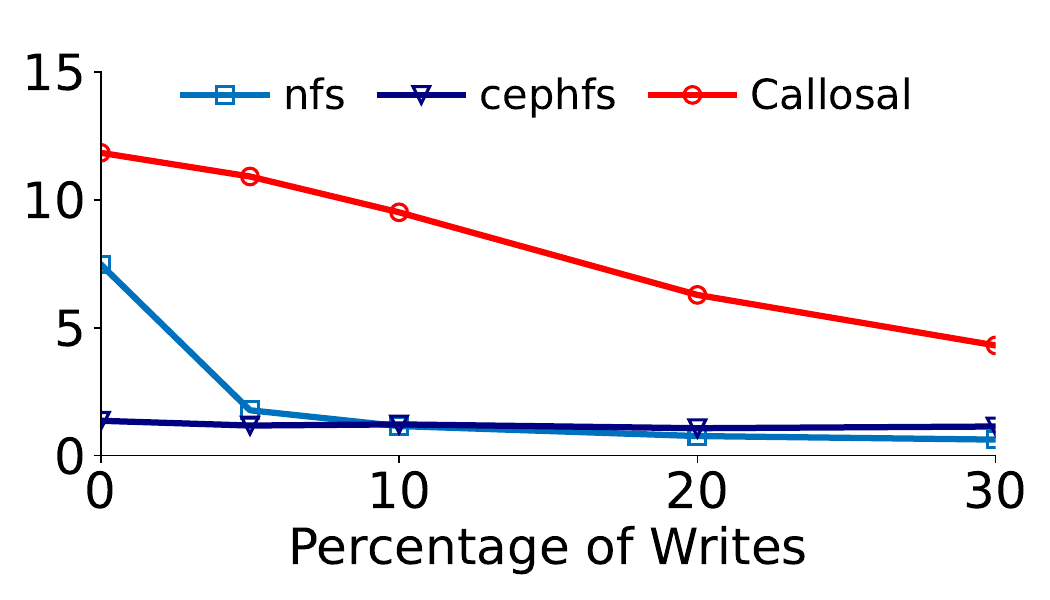}
\label{DRWOM_single_node}}
\\
\vspace{-1mm}
\quad \hspace{-8mm}
\subfigure[DRWOH 5$\%$ Writes]{%
\includegraphics[scale = .3225]{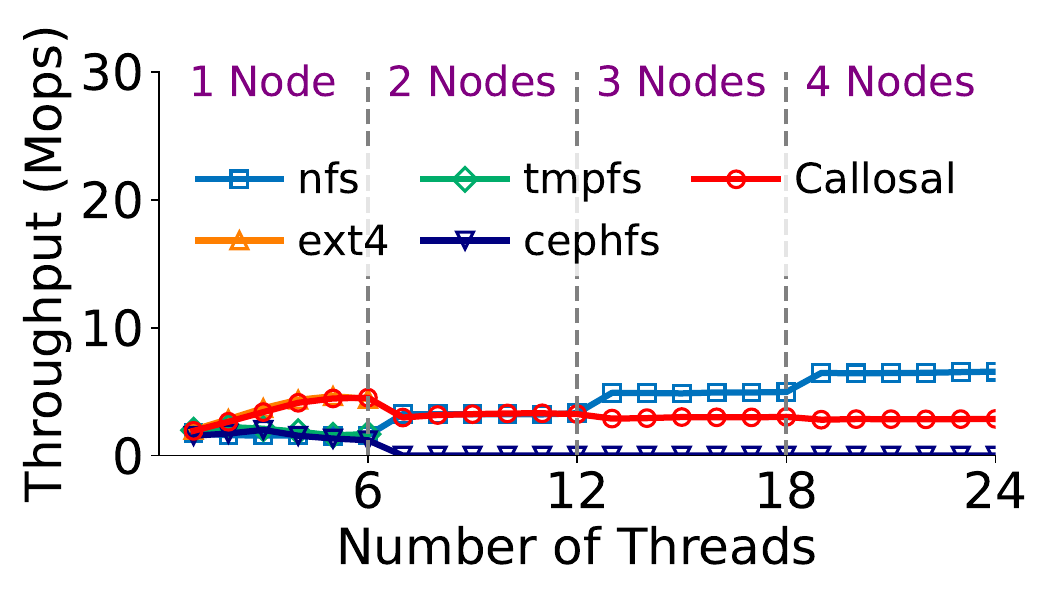}
\label{DRWOH}}
\subfigure[DRWOH]{%
\includegraphics[scale = .3225]{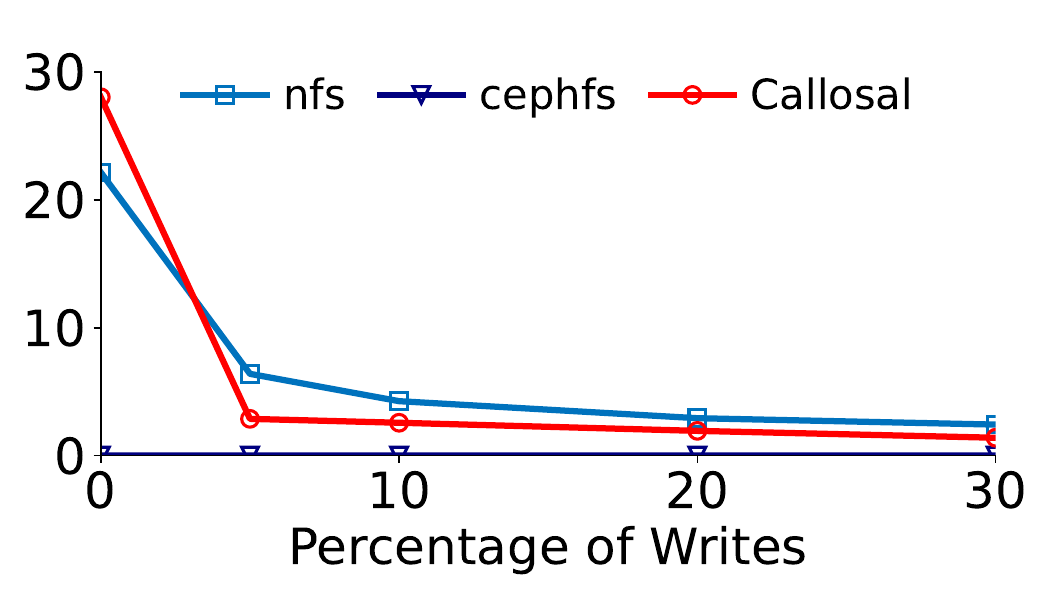}
\label{DRWOH_writes}}
\subfigure[DRWOH Single Node]{%
\includegraphics[scale = .3225]{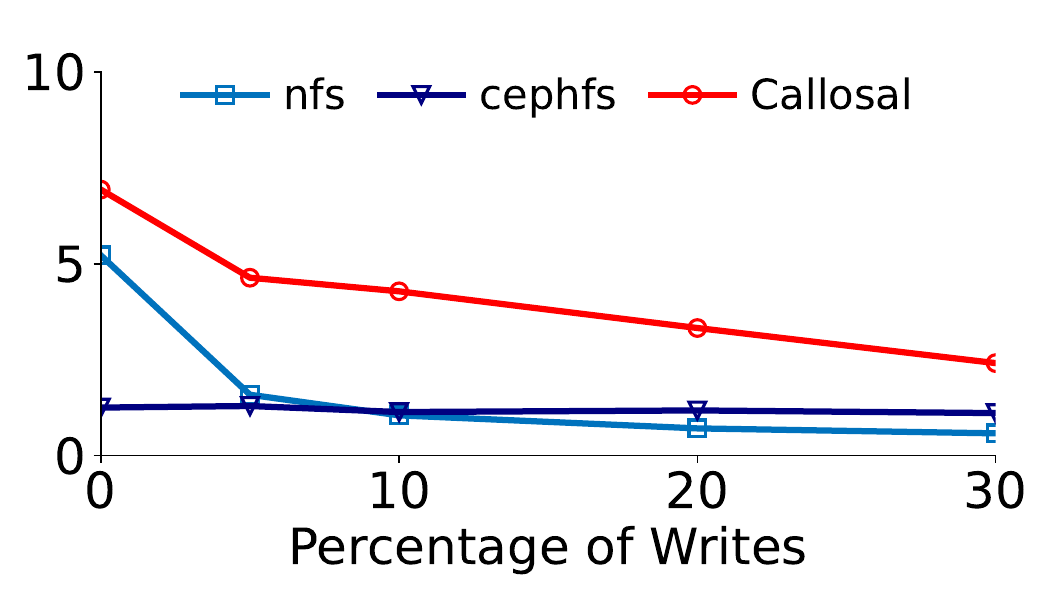}
\label{DRWOH_single_node}}
\vspace{-4mm}
\caption{Mixed read/write performance of a block in these scenarios. (a-c) DRWOM: reads or writes a private block in a shared file, and (d-f) DRWOH: reads or writes a shared block in a shared file. For DRWOM, \name outperforms NFS by 4.6$\times$ on average for 10-30$\%$ writes.
}
\vspace{-5mm}
\end{figure*}

\subsection{Write Throughput}

\paragraphb{DWOL} Figure~\ref{dwol} shows the results for DWOL in which every thread overwrites a block in a private file. 
All of the filesystems except CephFS scale linearly within a single node as expected since most coordination by each thread is local for inode metadata like \inode{}'s last modified time.
Here, NFS outperforms \fs{} somewhat as its RPC-based implementation effectively serializes those changes on a single node instead of relying on cross-node coherence messages to implement locking. However, \fs{}'s writes and metadata operations require additional state synchronization, which 
gives it the strong, external consistency guarantees that NFS does not provide, but this comes at the cost of some performance. In NFS, any readers added to the mix frequently read stale values.
CephFS provides strong consistency here, but it does not scale linearly across nodes, similar to DRBL.

\paragraphb{DWOM} Figure~\ref{dwom} shows the throughput of DWOM in which every thread overwrites a private block in a shared file. Linux serializes concurrent writes to each
file by holding a lock on \texttt{inode->i\_rwsem}, so throughput cannot scale within a node. NFS and \fs{} both maintain an independent \inode{} per client, and \fs{} only acquires the remote inode lock in write mode for size-adjusting writes. This allows concurrent writes when they originate on different nodes targeting separate pages and greatly reduces overhead (\S\ref{sec:consistency-model}).  NFS continues to outperform \fs{} for pure write-only workloads for the same reasons as with DWOL, though it gives up strong consistency. 
CephFS provides strong consistency, but since multiple clients write the same file, its less efficient approach to coherence causes its performance to collapse as the workload scales across nodes.

\paragraphb{DWOH} Figure~\ref{dwoh} shows the performance of DWOH in which all threads on all nodes repeatedly overwrite the same block in one shared file. Inherently, it is not possible to scale this workload while providing strong consistency. On a single node, \fs{} matches the performance of ext4 and NFS, but when scaling across nodes, the added cost of globally locking, invalidating, and fetching the up-to-date copy of the page increases dramatically, resulting in consistent but low throughput with \fs{} performing 136~Kops across four nodes.
Since NFS does not bother to serialize conflicting writes, it still achieves 3.2~Mops, making it difficult to see that \fs{}
 is still outperforming CephFS by up to 2 orders of magnitude,
benefiting again from its significantly lower overhead page-centric approach to
coherence.

\subsection{Mixed Read/Write Throughput}

\paragraphb{DRWOM} Read-only and write-only workloads show the limits of \fs{} performance, but its practical use is when different clients mix reads and writes since it provides strong consistency for reads.
To demonstrate these cases, we developed benchmarks by combining the existing \texttt{fxmark} benchmarks.
We call the first DRWOM; in this benchmark, each thread reads or writes a private block in a shared file; the fraction of operations that are reads or writes is configurable. Figure~\ref{DRWOM} shows DRWOM throughput with 5\% writes. Here, \fs{} vastly outperforms both NFS and CephFS by synchronizing cross-node writes at page granularity.

Figure~\ref{DRWOM_writes} varies the fraction of operations that are writes when run across 24 threads spanning all four nodes. Here, we can see that \fs{} outperforms NFS for all read/write ratios, even though NFS does not provide strong consistency. \fs{} performs about 
15.6~Mops with 30\%~writes.
Repeating the same comparison on a single node with 6~threads gives similar results (Figure~\ref{DRWOM_single_node}); however, running on a single node improves CephFS's performance to about 1.2~Mops regardless of the read/write ratio, causing it to outperform NFS for write-intensive workloads.

\paragraphb{DRWOH} Finally, we introduce a benchmark that we refer to as DRWOH. It is a high-contention benchmark where each thread reads or writes one block in one shared file, and the fraction of reads and writes varies. Figure~\ref{DRWOH} shows DRWOH throughput with 5\% writes.
Within a single machine, this is a high but tolerable cost; \fs{} matches ext4's performance within a node. But, as the workload scales across nodes, locking the page requires remote operations, which increases the duration of the write lock, causing a drop as the workload scales. Interestingly, within a single node, NFS performs more poorly than \fs{}. 
Across nodes, NFS outperforms \fs{}, but readers risk seeing stale values due to its weak consistency guarantees.

Figure~\ref{DRWOH_writes} shows the throughput of DRWOH for different write percentages when run on 24 threads spanning four nodes. With 0\% writes (100\% reads), \fs{} is slightly faster than NFS; however, NFS's weak consistency effectively allows it to avoid paying the full remote cost of synchronizing each operation, allowing it to outperform \fs{} when reads and writes are intermixed on highly contented pages. Figure~\ref{DRWOH_single_node} shows the same results on six threads on one node; within a single node \fs{} outperforms NFS. CephFS shows the same effect as with DRWOM, outperforming NFS as writes increase.

\subsection{Producer/Consumer Queue Application}

\begin{table}[t]
    \centering
    \small
    \begin{tabular}{r|rr} 
        \toprule
        \textbf{Buffer Size} & \textbf{CallosalFS (ops)} & \textbf{CephFS (ops)} \\ 
        \midrule
        1      & 4,718 & 124 \\ 
        1,000  & 16,740 & 138 \\ 
        5,000  & 33,809 & 128\\ 
        \bottomrule
    \end{tabular}
    \vspace{1mm}
    \caption{Throughput between a producer and consumer application with different buffer sizes for CallosalFS and Ceph.}
    \label{tab:appperf}
    \vspace{-10mm}
\end{table}

We also compared against CephFS using a simple single-producer/single-consumer (SPSC) queue implemented as a bounded ring buffer queue stored within files.
One file stores the producer and consumer indices, and the other contains the ring buffer elements. 
Similar workloads are common in distributed systems for data passing.
For example, MapReduce uses similar patterns to pass data between map tasks and
reduce tasks~\cite{mapreduce}.
We forego batching; each read/write only pops/pushes a single element. 
We run this SPSC queue on \fs{} and CephFS; NFS requires file locking due to its weak guarantees, which results in unacceptable performance. 
We tested three different ring buffer sizes, one with size 1 (enforces fairness), 1,000 (about 1 page), and 5,000 (spanning multiple pages).

\fs{}'s performance increases as the ring buffer 
size increases, 
likely due to two reasons. The first is that multiple operations can finish before an invalidation is processed, amortizing coherence costs. The second is due to reduced contention as the buffer spans pages. 
These numbers are lower than the \texttt{fxmark} numbers since we count the items transferred, not the total filesystem operations. CephFS's much larger coherence granularity, additional overhead, and its more expensive communication for synchronization dominate, so its performance does not increase with increased buffer sizes, 
as shown in Table~\ref{tab:appperf}. 
CephFS is more than an order of magnitude slower than \fs{} in the best case, with the
gap increasing with the size of the buffer.

\begin{figure}[t] 
\centering
\quad \hspace{-7mm}
\includegraphics[scale = .23]{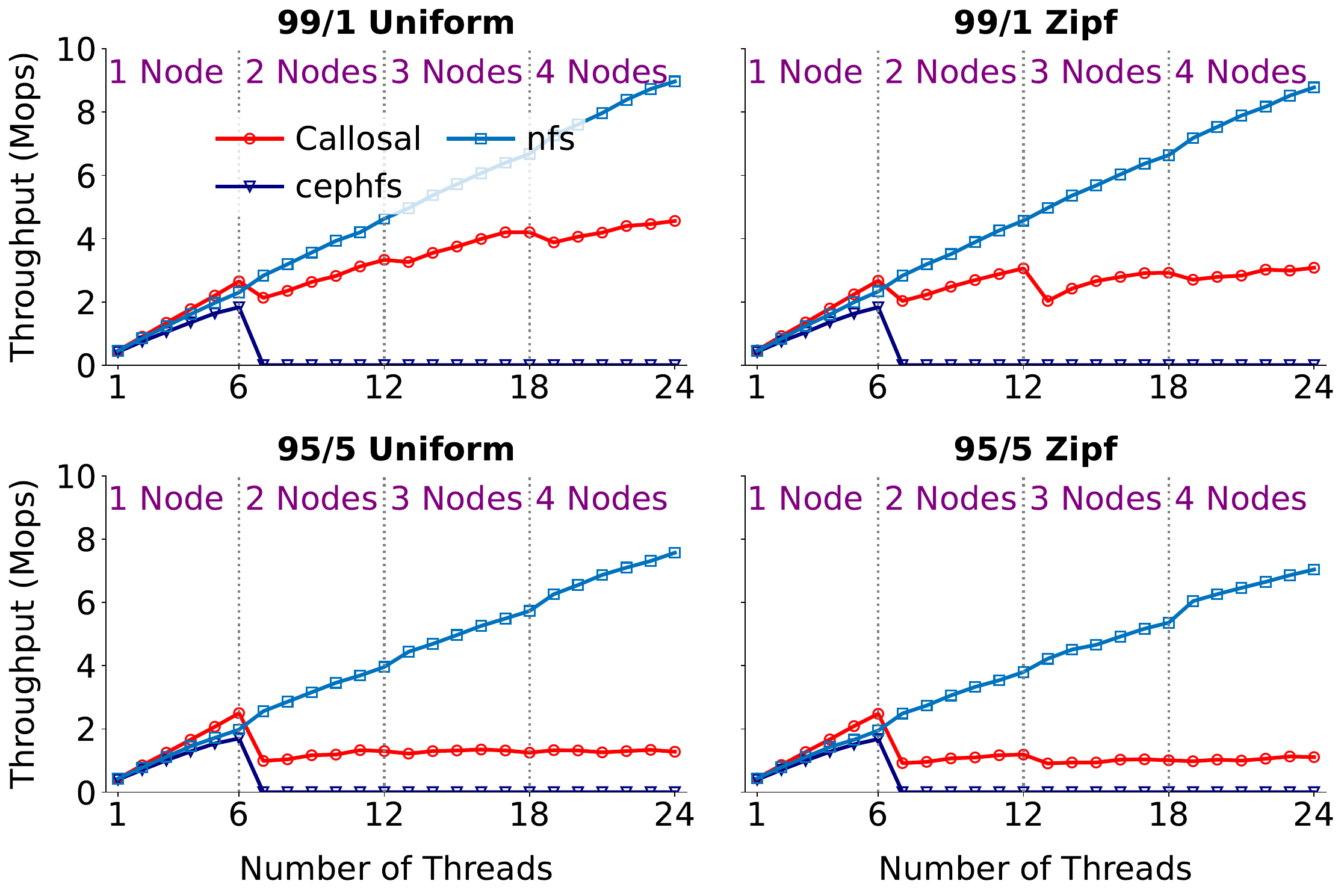}
\vspace{-7mm}
\caption{Key-value store throughput for 99/1 and 95/5 read/write workloads
under uniform and Zipfian access distributions. NFS scales but does not
provide correct key-value store behavior.}
\label{fig:dfs_kv}
\vspace{-6mm}
\end{figure}

\subsection{Key-Value Store Application}
We use a file-backed key-value store to evaluate \fs{} on an application that
depends on strongly consistent reads and writes to mutable state. The key-value
store uses ordinary file operations; it stores data in files shared by all
nodes, relies on the filesystem to maintain coherent cached pages, and 
does not implement a separate distributed coherence protocol in the application. Hence,
it shows that applications can  seamlessly share mutable file-backed state  through the
Linux filesystem interface while maintaining strong consistency.

The key-value store divides keys across worker threads with each key assigned
to a single writer and any worker allowed to read any key. This ownership rule
avoids write/write conflicts for individual keys, but reads still move data
across nodes when they target keys owned by remote workers. Readers are
synchronized safely with writers using a \texttt{seqlock}-like approach: writers
update item versions before and after an update; readers read optimistically and
validate that the read was atomic before using the result. This workload has
mostly local work with a small amount of cross-node mutation that must still be
observed consistently by readers. 
\fs{} and CephFS provide strong consistency for correct key-value store
behavior on this workload. NFS's weak consistency means that many operations
return stale data.

Our key-value store contains about 240,000 keys, resulting in a 75\% load factor.
Each entry has an 8\,B key, an 8\,B version, and a 128\,B value. The benchmark
issues ordinary \texttt{read()} and \texttt{write()} system calls. We run
uniform and $\theta=0.99$ Zipfian request distributions.

Figure~\ref{fig:dfs_kv} shows key-value store throughput as the workload is
scaled across threads and nodes. With 99\% gets and 1\% puts on both uniform
(upper-left) and Zipfian (upper-right) distributions, \fs{} continues to scale
after the workload spans nodes, reaching about 4.5\,Mops and 3\,Mops, respectively,
at four nodes. For 1\% writes under uniform access, 
this is 1.72$\times$ the single-node throughput. 
NFS's higher throughput is a byproduct of it producing incorrect, stale values on queries. 
Throughput collapses when using CephFS once the
workload crosses a single node. The 95\% get 5\% put workload shows the limit
of \fs{}'s design. Ownership transfers and invalidations limit \fs{}, and it
does not scale beyond a single node.
This key-value store
demonstrates that applications can store shared mutable state in ordinary files
and rely on the filesystem to keep cached reads and writes coherent across
nodes. Compared to NFS, \fs{} pays coherence costs but preserves correctness.
Compared to CephFS, \fs{} keeps coherence at page granularity and avoids the
large coordination overheads that dominate this workload.

\figsqueeze{-1mm}
\section{Related Work}
\label{sec:related}

\paragraphb{Memory and Disaggregation}
Many seminal works of the 1980s and 1990s developed efficient DSMs~\cite{li1986shared,ivy88}, including follow-on work that incorporated semantic knowledge of application objects to accelerate operations on objects like application-level locks and to reduce false sharing~\cite{treadmarks1994}. \name{} uses similar techniques e.g.\ to optimize updates to \inode{} metadata stored in \name{}'s DSM.

MIND~\cite{mind} and Concordia~\cite{concordia} proposed DSMs that use in-network cache coherence to provide efficient shared memory access across multiple nodes in a cluster.
This lets a coherence directory at the switch route page requests to the node(s) that currently hold them, reducing the cost of fetching pages by a round trip and invalidation costs via switch multicast.
\name{} uses these optimizations.

UniMem~\cite{zhong2024unimem} provides memory management in disaggregated data centers by creating a unified local-remote memory hierarchy providing seamless access to both local and remote memory resources. \name{} similarly combines page coherence via DSM with Linux's standard local page cache at each node.

\paragraphb{Low-latency Distributed Filesystems}
Recently, distributed file systems have been developed for low-latency storage.
Assise uses RDMA and NVM to let applications directly access and replicate state from userspace~\cite{assise}; \name{} avoids NVM as it is not commercially available in quantity. Assise also breaks transparency by requiring a special userspace library. Assise's leases provide strong consistency but require more costly invalidations than \name{}, since Assise restricts sharing to the file granularity rather than to pages. LineFS~\cite{linefs} accelerates Assise using ``smart'' network cards, but the same comparison applies, and \name{} relies only on standard RDMA network cards without general-purpose CPU cores. Octopus~\cite{octopus} and Ethane~\cite{ethane} use RDMA to build high-performance, low-latency access to a shared filesystem, but, similar to Assise, they both require NVM; neither provides transparency as they operate at the user level.
Lustre~\cite{lustre} is a high-performance filesystem designed specifically for large-scale cluster computing environments. It excels at handling massive datasets and multiple users accessing and modifying them simultaneously while offering consistency for core data and metadata operations; however, it generally operates at coarser granularity (MBs rather than KBs).
DFUSE~\cite{li2025enabling}, built on FUSE~\cite{vangoor2017fuse}, expands on distributed filesystem designs in FUSE by integrating lease synchronization with its userspace daemon into the kernel driver. This helps userspace filesystem designers maintain consistency with kernel filesystem operations without having to write through all the operations
to the userspace daemon. \name{} is a purely kernel-space approach.

blk-switch is a Linux kernel storage stack architecture designed to achieve microsecond-scale latency and high throughput~\cite{hwang2021rearchitecting}. It treats the storage stack like a network switch, leveraging the multi-queue design of modern storage and network hardware.
Similarly, i10~\cite{i10} achieves RDMA-like performance over TCP without requiring specialized hardware or kernel modifications. i10 offloads much of the TCP processing to network cards, reducing overhead and improving performance.
Both of these systems address block storage, not distributed file systems.

WekaFS~\cite{Weka2020Architecture} is a proprietary filesystem that synchronizes at 4~KB granularity (like \name{}) and targets smaller files compared to Lustre~\cite{wekavslustre}. WekaFS also allows for a writeback cache configuration~\cite{wekawriteback}. While it can use RDMA for communication, it does not use an in-network coherence directory.

DPC~\cite{bergman2026dpc} is a distributed page cache over CXL that enforces a single-copy invariant, in which remote nodes access pages via hardware-coherent CXL memory mappings. In contrast, \name{} targets commodity RDMA and programmable switches available today and integrates directly into the Linux kernel's page cache rather than relying on a Virtiofs/FUSE substrate.

\figsqueeze{-1mm}
\section{Conclusion}

\name{} is a new kernel page cache approach designed for shared disaggregated storage. By using network-accelerated DSM, \name is able to provide a transparent, consistent, and scalable page cache for filesystems designed for disaggregated block storage. Our results show that when used with \fs{}, 
this implementation outperforms NFS by around 4.6$\times$ for mixed read/write benchmarks when contention is not extreme. It also outperforms CephFS by an order of magnitude on multiple workloads.
For a key-value store, \fs{} scales 1.72$\times$ from one to four nodes with 1\%
writes and uniform access without sacrificing consistency. 

\ifarxiv
% !TEX root = ../paper.tex
\section*{Acknowledgments}
This material is based upon work supported by the National Science Foundation under Grant No. CNS-2245999, CNS-2047220 and CNS-2112562. Any opinions, findings, and conclusions or recommendations expressed in this material are those of the authors and do not necessarily reflect the views of the National Science Foundation.
This work was also supported in part by VMware, Meta and NetApp.

\fi
\section*{AI Disclosure}

We used Codex and Claude for language editing; debugging experimental
scripts and benchmark harnesses; and assistance writing scripts for data
analysis. We also used these tools to generate code for the prototype
key-value store, its benchmark harness, and associated plotting scripts.
The key-value store's design, synchronization approach, and workload were
specified by the authors. The generated code was manually reviewed by the
authors. AI-assisted code generation was not used as part of the
underlying filesystem implementation. The authors take full responsibility
for the accuracy, integrity, and originality of the work.

{
\balance
\bibliographystyle{plain}
\bibliography{bib/abr-short,bib/paper}
}
% \begin{sloppypar}
%\appendix
%\input{tex/appendix}
% \end{sloppypar}
\end{document}